\documentclass[10pt,preprint]{aastex}
\usepackage[]{epsfig}
\newcommand       \Angstrom     {\,{\rm \AA}}          

\newcommand       \cm           {\,{\rm cm}}
\newcommand       \erg          {\,{\rm erg}}
\newcommand       \eV           {\,{\rm eV}\,}

\newcommand	  \rmH		{{\rm H}}
\newcommand       \HH           {{\rm H}_2}
\newcommand       \K            {\,{\rm K}}

\newcommand	  \indx         {m}
\newcommand       \indxim       {m^{\prime\prime}}
\newcommand       \indxre       {m^{\prime}}

\newcommand	  \s		{\,{\rm s}}
\newcommand       \sr           {\,{\rm sr}}

\newcommand       \simlt        {\lesssim}
\newcommand       \simgt        {\gtrsim}
\newcommand       \gtsim        {\gtrsim}
\newcommand       \ltsim        {\lesssim}
\newcommand       \etapl        {\eta_{\rm PL}}
\newcommand       \ratesi       {\gamma_{\rm Si}}

\newcommand       \sisun        {\left[{\rm Si/H}\right]_{\odot}}

\newcommand       \nuere        {\langle h\nu\rangle_{\rm ERE}}
\newcommand	  \um	        {\mu{\rm m}}
\newcommand       \zsi          {\rm Z_{Si}}

\newcommand{\figwidth}{11.6cm}
\newcommand{\figwidthb}{11.0cm}	%size for fig11, which has larger caption

\countdef\decade=200
\advance\decade by \year
\countdef\hours=201
\advance\hours by \time
\divide\hours by 60
\countdef\mins=202
\advance\mins by \hours
\multiply\mins by 60
\multiply\hours by 100
\countdef\miltime=203
\advance\miltime by \hours
\advance\miltime by \time
\advance\miltime by -\mins

\shorttitle{Silicon Nanoparticles}

\begin{document}

\title{
%------------- enable for labelling preprint ---------------------------
        \vspace*{-2.0em}
        {\normalsize\rm revised version, 
	submitted to {\it The Astrophysical Journal}}\\
        \vspace*{1.0em}
%-----------------------------------------------------------------------
	Are Silicon Nanoparticles an Interstellar Dust Component?%
% following is useful to identify draft version
%	\\{\small DRAFT: \today}%
	}

\author{Aigen Li and B.T. Draine}
\affil{Princeton University Observatory, Peyton Hall,
        Princeton, NJ 08544, USA;\\
        {\sf agli@astro.princeton.edu, draine@astro.princeton.edu}}

\begin{abstract}
Silicon nanoparticles (SNPs) have been proposed as the source of 
the observed ``extended red emission'' (ERE) from interstellar dust.
We calculate the thermal emission expected from such particles, both in a 
reflection nebula such as NGC 2023 and in the diffuse 
interstellar medium (ISM).
Pure neutral Si SNPs would emit at
16.4$\um$, while Si/SiO$_2$ SNPs (both neutral and charged)
produce a feature at 20$\um$.
Observational upper limits on the 16.4$\um$ and 20$\um$ features
in NGC 2023 impose
upper limits of $< 1.5$ppm in pure Si SNPs, and
or $< 0.2$ppm in Si/SiO$_2$ SNPs.
The observed ERE intensity from NGC 2023 then gives
a lower bound on the required photoluminescence efficiency $\etapl$.
For foreground extinction $A_{0.68\um} = 1.2$, 
we find $\etapl > 5\%$ for Si SNPs, or $\etapl > 24\%$
for Si/SiO$_2$ SNPs in NGC 2023.
Measurement of the R band extinction toward the ERE-emitting region could
strengthen these lower limits.
The ERE emissivity of the diffuse interstellar medium appears to require
$\gtsim$42\% ($\gtsim$33\%) of solar Si abundance in Si/SiO$_2$ (Si) SNPs.
We predict IR emission spectra and show that
DIRBE photometry appears to rule out such high abundances 
of free-flying SNPs in the diffuse ISM. 
If the ERE is due to SNPs, they must be 
either in clusters or attached to larger grains.
Future observations by SIRTF will be even more sensitive to the
presence of free-flying SNPs.

\end{abstract}

\keywords{dust, extinction --- infrared: ISM: lines and bands
--- reflection nebulae: NGC 2023}

\section{Introduction\label{sec:intro}}

First detected in the Red Rectangle (Schmidt, Cohen, \& Margon 
1980), ``extended red emission'' (ERE) from interstellar dust
consists of a broad,
featureless emission band 
between $\sim$5400$\Angstrom$ and 9000$\Angstrom$,
peaking at $6100\ltsim \lambda_{\rm p} \ltsim 8200\Angstrom$, and
with a width $600\Angstrom\ltsim {\rm FWHM}\ltsim 1000\Angstrom$.
The ERE has been seen in 
a wide variety of dusty environments: the diffuse interstellar
medium (ISM) of our Galaxy, reflection nebulae, planetary nebulae, 
HII regions, and other galaxies (see Witt, Gordon, \& Furton 1998 
for a summary). The ERE is generally attributed to 
photoluminescence (PL) by some component of interstellar grains, powered 
by ultraviolet (UV)/visible photons. The photon conversion efficiency of
the diffuse ISM has been determined to be near 
($10\pm 3$)\% (Gordon et al.\ 1998; Szomoru \& Guhathakurta 1998) assuming
that all UV/visible photons absorbed by interstellar grains are absorbed
by the ERE carrier.
The actual photoluminescence efficiency $\etapl$
of the ERE carrier
must exceed $\sim 10\%$, since the ERE carrier
cannot be the only UV/visible photon absorber.

Various forms of carbonaceous materials -- 
hydrogenated amorphous carbon (HAC) (Duley 1985; Witt \& Schild 1988), 
polycyclic aromatic hydrocarbons (PAHs) 
(d'Hendecourt et al.\ 1986)\footnote{%
    High photoluminescence efficiencies can be obtained by PAHs.
    Arguments against PAHs as ERE carriers include:
    (1) presence of
    sharp structures in the luminescence spectra of individual 
    PAH molecules in contrast to the featureless nature of the 
    interstellar ERE spectra;
    (2) lack of spatial correlation between the ERE 
    and the PAH IR emission bands in the compact HII region Sh 152
    (Darbon et al.\ 2000) and in the Orion Nebula (Perrin \& Sivan 1992);
    (3) ERE detection in the Bubble Nebula where no PAH emission has been
    detected (Sivan \& Perrin 1993);
    (4) nondetection of ERE emission in reflection nebulae 
    illuminated by stars with effective temperatures
    $T_{\star}<7000\K$ (Darbon, Perrin, \& Sivan 1999) whereas
    PAHs emission bands have been seen
    in such regions (e.g., see Uchida, Sellgren, \& Werner 1998) 
    and are expected for the PAH emission model 
    (A. Li \& B.T. Draine 2001, in preparation).   
    Argument (1) may not be fatal since 
    a featureless band may result from a mixture
    of many individual PAH molecules and ions.%
    }
quenched carbonaceous composite (QCC) (Sakata et al.\ 1992), 
C$_{60}$ (Webster 1993), 
coal (Papoular et al.\ 1996), 
PAH clusters (Allamandola, private communication), and 
carbon nanoparticles (Seahra \& Duley 1999)\footnote{%
	Seahra \& Duley (1999) argued that small carbon 
	clusters were able to meet both the ERE profile and the PL efficiency
	requirements. 
	However, this hypothesis appears to be ruled out by
	non-detection in NGC 7023 of the 1$\mu$m ERE peak (Gordon et al.\ 2000)
	predicted by the carbon nanoparticle model.} 
-- have been proposed as carriers of ERE. 
However, most candidates appear to be unable 
to simultaneously match the observed ERE spectra 
and the required PL efficiency (see Witt et al.\ 1998 for details).

Very recently, Witt et al.\ (1998) and Ledoux et al.\ (1998) 
suggested crystalline silicon nanoparticles (SNPs) with
15\AA\ -- 50\AA\ diameters as the carrier on the 
basis of experimental data showing that SNPs could provide a close 
match to the observed ERE spectra and satisfy the quantum efficiency 
requirement. It was estimated by Witt et al.\ (1998) and Ledoux et al.\ 
(1998) that SNPs account for $\ltsim 5\%$ of the total interstellar 
dust mass, with Si/H $\approx6$~ppm.
Zubko et al.\ (1999) modelled the interstellar 
extinction curve taking SNPs as an interstellar dust component 
containing
Si/H $\approx 18$~ppm,
or $\sim$50\% of the solar abundance $\sisun=36$~ppm.
More recently, Smith \& Witt (2001) have further developed the SNP
model for the ERE, concluding that the observed ERE in the diffuse ISM
can be explained with Si/H = 6~ppm in SiO$_2$-coated SNPs with Si core
radii $a\approx17.5\Angstrom$.

The purpose of this paper is to test the SNP hypothesis.
We calculate the IR spectra for SNPs in a reflection nebula
-- NGC 2023 -- and compare with observations.
We show that the SNPs in NGC 2023 contain Si/H $\ltsim 1.5$~ppm.
We re-estimate the {\it minimum} Si depletion in SNPs required to account
for the observed ERE intensity in the diffuse ISM, and calculate
the IR emission expected from such particles.
We show that existing DIRBE photometry appears to rule out the abundances 
of free-flying SNPs required
to account for the ERE emissivity of the diffuse ISM.
Future observations by the
{\it Space Infrared Telescope Facility} (SIRTF) will be even more
sensitive to the abundance of SNPs in the diffuse ISM.

In \S\ref{sec:physics} we discuss the optical properties of 
silicon nano-crystals and glassy SiO$_2$ (\S\ref{sec:opct})
as well as the heat capacities of pure Si and
Si core-SiO$_2$ mantle grains (\S\ref{sec:enthal}). 
In \S\ref{sec:results} we carry out calculations for
the IR emission spectra of SNPs (with or without SiO$_2$ coatings),
and discuss their implications. In \S\ref{sec:discussion} we
discuss the effects of grain shape, mantle thickness, and
sources of uncertainties. Our conclusions are presented 
in \S\ref{sec:summary}.
 
\section{Grain Physics\label{sec:physics}}

Experimental studies indicate that high photoluminescence efficiencies
are only observed from Si nanocrystals when their surfaces have
been ``passivated'' by oxidation or hydrogenation -- otherwise
electron-hole pairs recombine nonradiatively
(see, e.g., Kovalev et al. 1999).
Accordingly, SNPs with oxide coatings are likely to be of primary
interest as the source of the ERE.
However, we will also discuss the infrared emission of pure Si SNPs
in order to show that even if the Si core dominates the IR emission,
the IR spectral signature is still conspicuous.

\subsection{Optical Properties\label{sec:opct}}

The optical properties of crystalline Si nanoparticles 
are controversial. Some studies (see Yoffe 2001 for a recent 
review) have concluded that the optical properties of Si nanoparticles 
differ substantially from those of bulk Si.
This has been attributed to the quantum confinement effect 
(Wang \& Zunger 1994; Tsu, Babic, \& Ioriatti 1997).
Koshida et al.\ (1993; hereafter KKS93)
have published dielectric constants for crystalline nanosilicon which 
are considerably smaller than those of {\it bulk}-Si. 
We find that the absorption and reflectivity measurements for 
porous silicon (e.g., Koshida et al.\ 1993, de Filippo et al.\ 2000)
appear to be inconsistent with the KKS93 optical constants;
instead, they can be approximately reproduced using a mixture of
Si and voids, if the Si component
is described using the optical constants of bulk Si,
as previously found by
Kovalev et al.\ (1996), 
Theiss (1997),
L\'erondel, Mad\'eore, \& Muller (2000), and 
Diesinger, Bsiesy, \& H\'erino (2001).
Oxide layers are generally also present in laboratory samples of porous Si;
we find that the absorption and reflectance spectra can also be 
satisfactorily reproduced
using Bruggeman effective medium theory (Bohren \& Huffman 1984)
for a mixture of voids, SiO$_2$, and bulk-Si.

We take the following ``synthetic'' approach to obtain the 
complex refractive index
$\indx(\lambda)=\indxre+i\indxim$ for bulk Si at low temperatures,
with no electron-hole pairs present (the contribution of electron-hole
pairs will be treated separately).
For $0.01 < \lambda < 1.2\um$ 
we take $\indxim$ of Adachi (1999) for bulk crystalline Si.
For $1.2 < \lambda < 3.6\um$ we set 
$\indxim=2.4\times 10^{-8}$ 
estimated from the (room temperature) absorption coefficient 
$\alpha \approx 10^{-3}\cm^{-1}$ of
bulk Si between 1 and 3$\um$ (Gray 1972).
For $3.6 < \lambda < 25\um$ we take those of
Palik (1985) for crystalline bulk Si; extrapolation is then made
for ${\rm \lambda > 25\mu m}$.\footnote{%
	We do not adopt those of Palik (1985)
	for $\lambda > 25\um$ since they were measured at $T\approx300\K$,
	and the $\lambda > 25\um$ absorption is dominated by
	thermally-excited electron-hole pairs.
	In the absence of such pairs, we approximate
	$\indxim(\lambda) \approx \indxim(25\um)(25\um/\lambda)$.
	The IR bands of Si nanoparticles may become stronger
	due to the symmetry-breaking surfaces 
        (e.g., see Hofmeister, Rosen, \& Speck 2000). 
	New IR bands, forbidden in bulk Si, may also appear
	(Adachi, private communication).
	Unfortunately, IR measurements of Si nanoparticles are unavailable,
	so we must use the optical properties of bulk Si.
	}
After smoothly joining the adopted $\indxim$, we calculate 
$\indxre$ from $\indxim$ 
through the Kramers-Kronig relation (Bohren \& Huffman 1983, p.\ 28).

The Si long-wavelength absorption properties will depend on 
whether any free electrons or holes are present.
SNPs are thought to luminesce only when uncharged and containing exactly
one electron-hole pair, which recombines radiatively in
$\sim10^{-3}\s$  (Smith \& Witt 2001).
The thermal energy remaining after luminescence will then be radiated by
a SNP with no free electrons or holes, at least in the diffuse ISM where
the cooling time is short compared to the time between absorption of UV
photons.
For these SNPs, we use the optical properties of bulk Si.

Charged SNPs will contain at least one free electron or hole, which we
assume contributes to the dielectric function
\begin{equation}
\delta\epsilon \approx 
\frac{-(\omega_p\tau)^2}{(\omega_p\tau)^2 + i\omega\tau}
\end{equation}
\begin{equation}
\omega_p^2 = \frac{3|Z|}{4\pi a_{\rm Si}^3}\frac{e^2}{4\pi m_{\rm eff}}
\end{equation}
where we adopt an effective mass $m_{\rm eff}\approx 0.2 m_e$
for electrons or holes in Si.\footnote{%
	For a positively charged SNP, the relevant mass is the effective
	mass of a hole.
	There are two degenerate valence band maxima, with hole
	masses $0.49m_e$ and $0.16m_e$ (Ashcroft \& Mermin 1976, p. 569);
	we adopt $0.2m_e$ as a representative value.
	The conduction band minima are characterized by electron effective
	masses of 1.0 and 0.2$m_e$; these would be relevant for 
	negatively-charged SNPs.
	}
Scattering by the Si boundary will result in 
$\tau\approx a/v_F$, where we take $v_F\approx 10^8\cm\s^{-1}$ 
as the typical velocity\footnote{%
   Our estimate of $\tau$ is by no means exact,
   but the value of $\tau$ is unimportant in the case of oxide-coated SNPs,
   since the oxide layer dominates the IR emission.
   As mentioned above, interstellar SNPs are thought to be oxide-coated.
   }

\begin{figure}[h]
\begin{center}
\epsfig{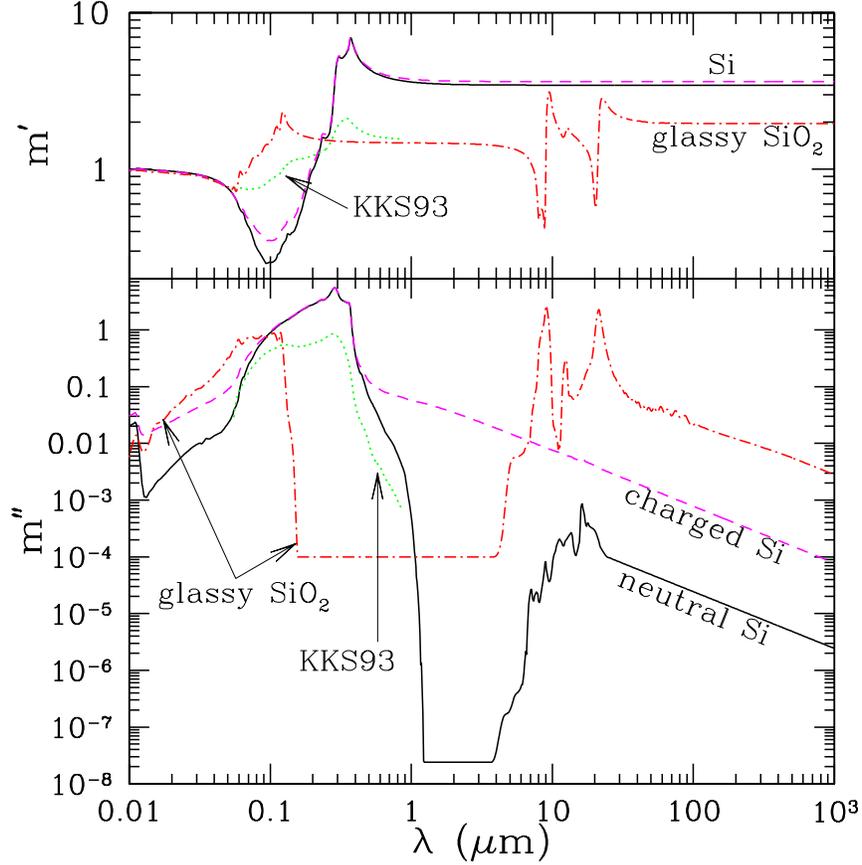}
\end{center}\vspace*{-2em}
\caption{
        \label{fig:opct}
        \footnotesize
        Optical constants $\indxre$ (upper panel), $\indxim$ (lower panel) 
        of neutral Si (solid lines), 
        charged Si ($Z=+1$, $a=10\Angstrom$; dashed lines),
        and SiO$_2$ glass (dot-dashed lines). Also plotted are 
        those of Koshida et al.\ 1993 (dotted lines; labelled by KKS93).
        }
\end{figure}

In Figure \ref{fig:opct} we plot the resulting
optical constants for crystalline SNPs, both for neutral SNPs and
SNPs with a radius $a=10\Angstrom$ and $Z=+1$.
We see that the hole dominates the absorption for $\lambda \gtsim 1 \micron$,
and completely overwhelms the weak vibrational bands in the
7--25$\micron$ region which provide the infrared absorption in neutral Si.
For comparison, we also show the optical constants from KKS93.

For glassy SiO$_2$, we take $\indxim$ from Palik (1985) 
for $0.01 < \lambda < 7\um$ but with one modification: 
for $0.15 < \lambda < 3.6\um$ 
we set $\indxim=1.0\times 10^{-4}$ 
based on the absorption coefficient measured 
by Harrington et al.\ (1978). 
For $7 < \lambda < 500\um$ we take $\indxim$ from 
Henning \& Mutschke (1997) for glassy SiO$_2$ sample at 100$\K$
(the optical properties of glassy SiO$_2$ are not sensitive to
temperature, see Henning \& Mutschke 1997). 
For $\lambda >500 \um$, we approximate
$\indxim(\lambda) \approx \indxim(500\um)(500\um/\lambda)$.
Again, the real part $\indxre$ is calculated from 
the Kramers-Kronig relation. 
The results are also presented in Figure \ref{fig:opct}.

\begin{figure}[h]
\begin{center}
\epsfig{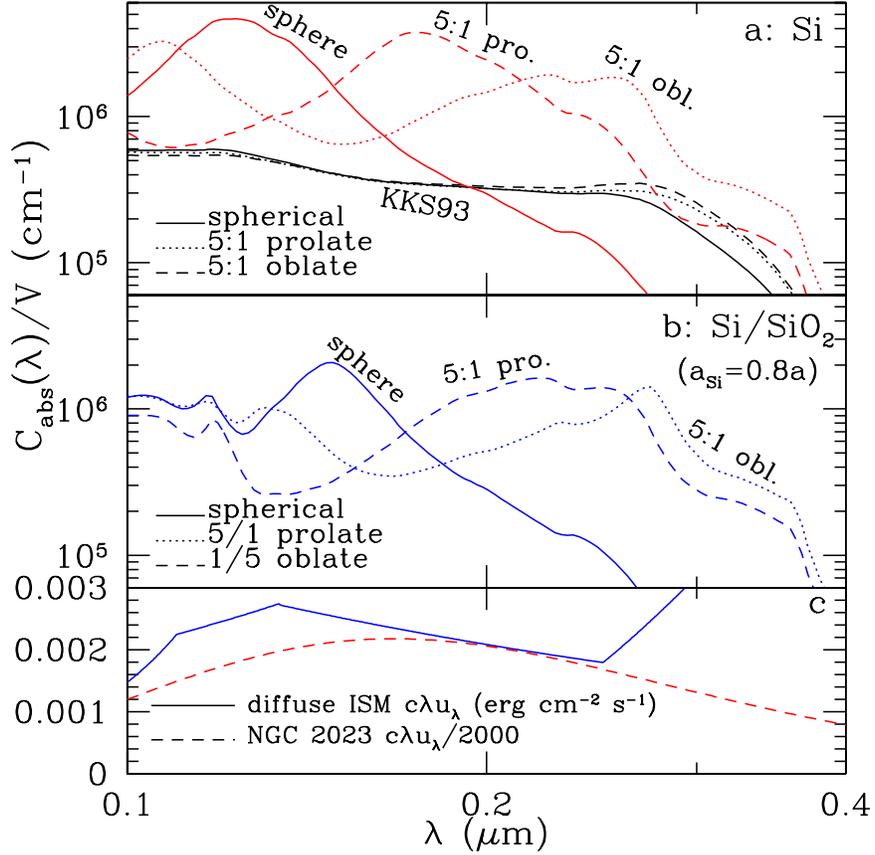}
\end{center}\vspace*{-2em}
\caption{
        \label{fig:cabs}
        \footnotesize
	(a): Absorption cross sections per unit volume 
        of spherical Si grains (solid line),
        5:1 prolate Si grains (dashed line),
        5:1 oblate Si grains (dotted line).
        Also plotted are the spherical, 5:1 prolate, 5:1 oblate
        grains using the optical constants of 
        Koshida et al.\ (1993) (labelled by KKS93).  
	(b) Absorption cross sections per unit volume 
        of spherical Si/SiO$_2$ grains (solid line),
        5:1 prolate Si/SiO$_2$ grains (dashed line),
        5:1 oblate Si/SiO$_2$ grains (dotted line).
	All grains are taken to have the same Si core volume
        fraction 51\% (for the spherical case, $a_{\rm Si}=0.8a$).
        (c): The solar neighbourhood interstellar radiation field 
        (Mathis, Mezger, \& Panagia 1983) 
        and the NGC 2023 radiation field at a position 
        60$^{\prime\prime}$ S of HD 37903 (see \S\ref{sec:ngc2023}).
        }
\end{figure}

In Figure \ref{fig:cabs} we show
the absorption cross sections per unit volume calculated for
both spherical and spheroidal
(a) pure Si grains, and
(b) Si/SiO$_2$ grains, with the Si in a confocal spheroidal core 
containing a fraction $(0.8)^3=51\%$ of the volume.
We also plot in Figure \ref{fig:cabs}c
the radiation fields for the diffuse ISM and for NGC 2023 
(60$^{\prime\prime}$ S of HD 37903; see \S\ref{sec:ngc2023}). 

Figures \ref{fig:cabs}a and \ref{fig:cabs}b clearly show 
that in the wavelength range 
where Si or Si/SiO$_2$ grains absorb most ($\lambda \simlt 0.25\mu$m)
the grain shape can substantially affect the wavelength-dependence
of the ultraviolet absorption.
However, we will see below (\S\ref{sec:discussion}) that the
integrated energy absorption rate, 
and the resulting emission spectrum, 
is surprisingly insensitive to grain shape.

\subsection{Enthalpies\label{sec:enthal}}

The experimental specific heat of bulk crystalline Si 
can be approximated by a Debye model (with dimensionality $n=3$)
and Debye temperature $\Theta=530$~K, while SiO$_2$ glass can be 
approximated by a model where 1/3 of the vibrational modes are
distributed according to a Debye model with $\Theta=275$~K, 
and 2/3 of the modes according to a Debye model with $\Theta=1200$~K.

For very small particles at low temperatures, the discrete 
nature of the vibrational spectrum becomes important.
Let $N_{\rm atom}^{\rm Si}$ and $N_{\rm atom}^{\rm SiO_2}$ 
be the number of atoms in the Si core and SiO$_2$ mantle, and 
$N_{\rm atom}\equiv N_{\rm atom}^{\rm Si} + N_{\rm atom}^{\rm SiO_2}$
be the total number of atoms.
For pure Si clusters, we set $N_m=3N_{\rm atom}^{\rm Si}-6$ vibrational 
modes where the ``$-6$'' term allows for the translational and rotational 
degrees of freedom since the energy from photon absorption is only
distributed among the vibrational modes.
For Si/SiO$_2$ nanoparticles, we assume $N_m=3N_{\rm atom}^{\rm Si}-3$ 
modes distributed according to a $\Theta=530$~K Debye model,
$N_m=2N_{\rm atom}^{\rm SiO_2}-2$ modes distributed 
according to a $\Theta=1200$~K Debye model, 
and $N_m=N_{\rm atom}^{\rm SiO_2}-1$ modes
distributed according to a $\Theta=275$~K Debye model.
We assume the mode frequencies to be distributed following eqs.(4-6) 
and eq.(11) of Draine \& Li (2001). The specific heat is calculated
treating the modes as harmonic oscillators, with the continuum limit used
for large particles and high energies (see Draine \& Li 2001). 

\section{IR Emission Spectrum and Its Implications\label{sec:results}}

Let $\zsi$ be the amount of Si in the grains (by number)
relative to total hydrogen.
The IR emissivity per unit solid angle per H nucleon 
from the SNPs is
\begin{equation}\label{eq:j_lambda}
j_{\lambda} (a) =
\frac{\zsi}{N_{\rm Si}(a)}
C_{\rm abs}(a,\lambda)
\int^{\infty}_{0} dT\ B_{\lambda}(T)\ P(a,T)
\end{equation}
where 
$N_{\rm Si}(a)$ is the number of Si atoms in a grain of radius
$a$
[$N_{\rm Si} = (\pi a^3/21 m_\rmH)\rho_{\rm Si}$ for pure Si nanoparticles,
and
$N_{\rm Si} = \pi a_{\rm Si}^3\rho_{\rm Si}/21 m_\rmH +
\pi (a^3-a_{\rm Si}^3)\rho_{\rm SiO_2}/45m_\rmH$ for grains with
Si cores of radius $a_{\rm Si}$ and SiO$_2$ mantles, where
$\rho_{\rm Si}$, $\rho_{\rm SiO_2}$ are the mass densities of crystalline 
Si ($\approx 2.42$\,g\,cm$^{-3}$) and glassy SiO$_2$ 
($\approx 2.2$\,g\,cm$^{-3}$), respectively, and
$m_{\rm H}$ is the mass of a hydrogen atom]; 
$C_{\rm abs}(a,\lambda)$ is the absorption 
cross section of the spheroidal core-mantle grain at wavelength
$\lambda$, calculated using the optical
constants discussed in \S\ref{sec:opct}; 
$B_{\lambda}(T)$ is the Planck function;
$P(a,T)dT$ is the probability that the grain temperature will
be in $[T,T+dT]$. For a given radiation field, we calculate $P(a,T)$ for 
small grains employing the ``thermal-discrete'' method 
(Draine \& Li 2001).\footnote{%
	To be precise, the energy released 
	in the form of PL is not available as heat.
        In NGC 2023, with a mean energy
	$\langle h\nu\rangle_{\rm ERE}\approx 1.8\eV$ for ERE photons,
        the fraction of absorbed photon energy lost to photoluminescence 
        $\etapl\langle h\nu\rangle_{\rm ERE}/\langle h\nu \rangle_{\rm abs}$
	is only $0.21\etapl$ for pure Si 
	($\langle h\nu\rangle_{\rm abs}=8.6\eV$)
	or $0.19\etapl$ for SiO$_2$-coated Si 
	($\langle h\nu\rangle_{\rm abs}=9.3\eV$).
	Even for $\etapl\rightarrow 1$, 
	this is a minor correction which does not
	alter our conclusions.}
We characterize the intensity of the illuminating starlight by $\chi$,
the intensity at 1000$\Angstrom$ relative to the Habing (1968) radiation
field.
For NGC 2023 the spectrum is assumed to be a $22000\K$ dilute blackbody
cutoff at the Lyman edge; for the diffuse interstellar medium
we take the spectrum of Mathis, Mezger, \& Panagia (1983), with
$\chi=1.23$.

\subsection{NGC 2023\label{sec:ngc2023}}

The reflection nebula NGC 2023, at a distance $D=450$pc, 
is illuminated by the B1.5V star HD 37903
($T_{\rm eff}=22,000$K, $L_{\star}=7600 L_{\odot}$).
A modest HII region surrounds the star, 
beyond which is a photodissociation region
at an estimated distance
$\sim 5\times10^{17}\cm$ from the star.
The radiation field is expected to have an intensity $\chi\approx 5000$
at this distance from the star.
While $\chi\approx5000$ is consistent with models to reproduce the
observed IR and far-red
emission from UV-pumped
$\HH$ in the bright ``emission bar'' 80$^{\prime\prime}$ S of HD 37903 
(Draine \& Bertoldi 1996, 2000), the integrated infrared
5--60$\um$ surface brightness at a position 60$^{\prime\prime}$ S of
HD 37903 (Figure \ref{fig:emsn_ngc2023_si}) points to a lower value of $\chi$.
We will assume that the dust producing the measured infrared
spectrum and ERE is illuminated by a radiation field with 
$\chi\approx 2000$.\footnote{%
	As explained in \S\ref{sec:discussion}, our analysis is insensitive
	to the precise value assumed for $\chi$.}

\begin{figure}[h]
\begin{center}
\epsfig{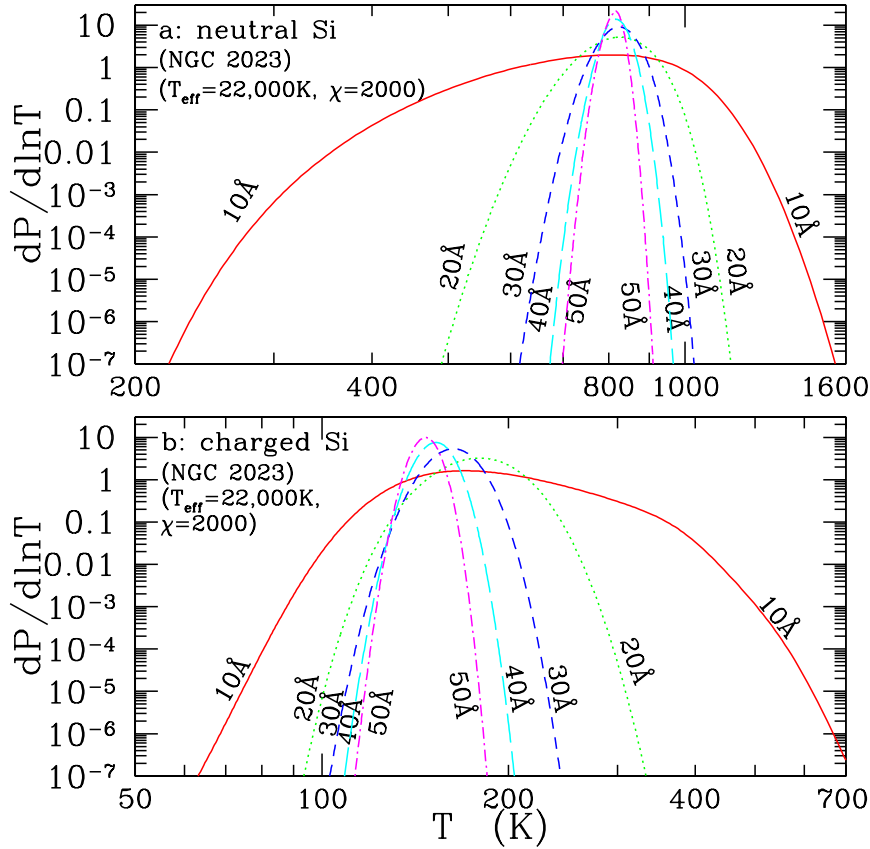}
\end{center}\vspace*{-2em}
\caption{
        \footnotesize
        \label{fig:PTdngc2023_si}
        Temperature distribution functions 
        for pure Si spheres of various radii,
	using the optical properties of (a) neutral particles and
	(b) particles with charge $Z=+1$,
	exposed to the radiation field in NGC 2023.
        Curves are labelled by grain radius $a$.
	}
\end{figure}

\begin{figure}[h]
\begin{center}
\epsfig{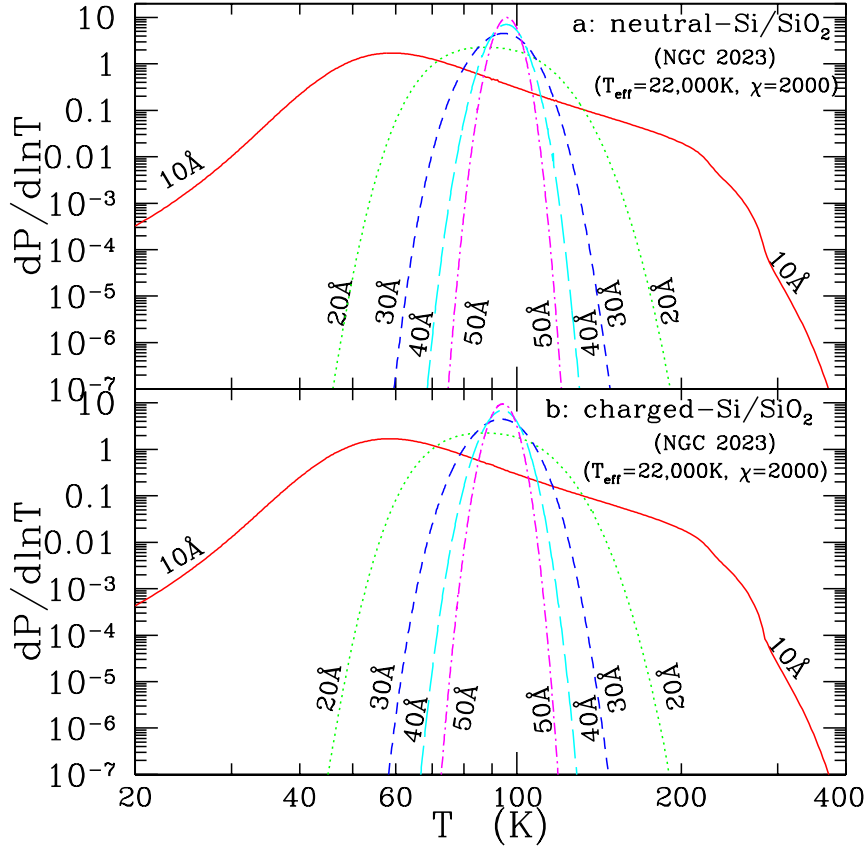}
\end{center}\vspace*{-2em}
\caption{
	\footnotesize
	\label{fig:PTdngc2023_sisio2}
	Temperature distribution functions for Si/SiO$_2$ spheres
	of various radii, with the Si cores assumed to be (a) neutral and
	(b) charged with $Z=+1$,
	for grains exposed to the radiation field in NGC 2023.
        Curves are labelled by the Si core radius $a_{\rm Si}$
        ($\approx 0.8 a$ where $a$ is the Si/SiO$_2$ radius).
        }
\end{figure}

\begin{figure}[h]
\begin{center}
\epsfig{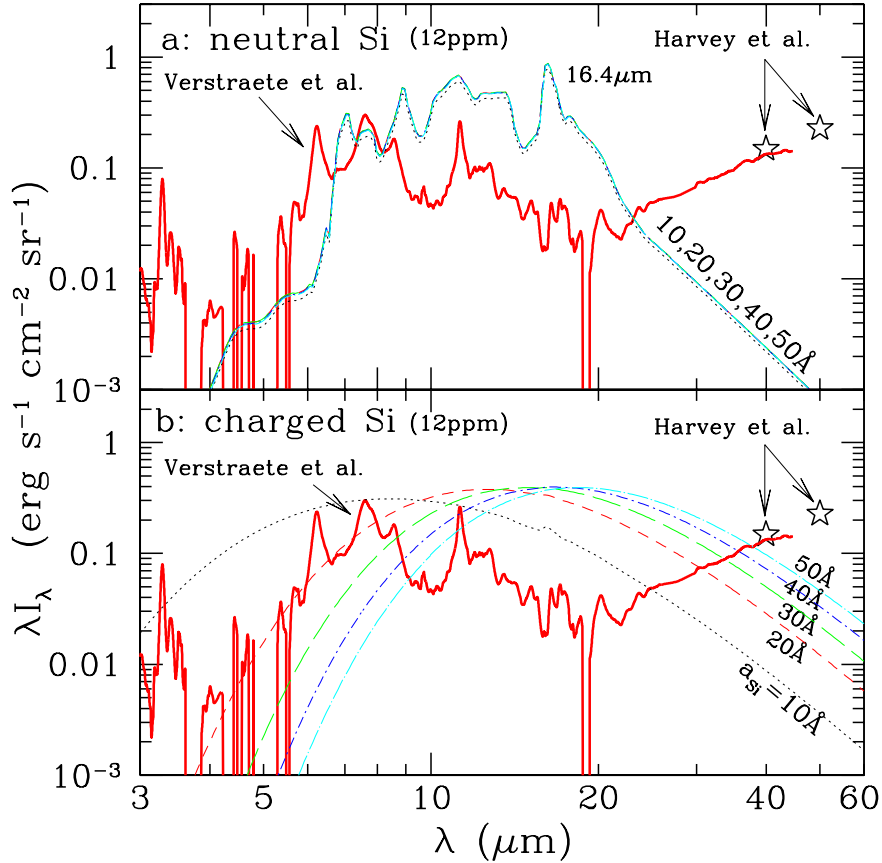}
\end{center}\vspace*{-2em}
\caption{
	\label{fig:emsn_ngc2023_si}
	\footnotesize
	Predicted IR emission spectrum for Si/H=12~ppm in
	(a) neutral and (b) $Z=+1$
	pure Si nanoparticles in NGC 2023.
	For the neutral Si nanoparticles, the emission is nearly
	independent of particle size; for the charged Si nanoparticles,
	spectra are labelled by grain radius.
	Heavy solid line:
	ISO SWS spectrum (Verstraete et al.\ 2001).
        Stars: KAO photometry (Harvey et al.\ 1980).
	For the 16.4$\micron$ feature not to exceed the ISO spectrum,
	the upper limit on neutral Si SNPs is Si/H $\ltsim 0.8$~ppm.
	The upper limit on charged Si SNPs alone is Si/H $\ltsim 1.5$~ppm.
	The upper limit on Si SNPs (both neutral and charged)
	is Si/H $\ltsim 1.5$~ppm.
	}
\end{figure}

In Figures \ref{fig:PTdngc2023_si} and \ref{fig:PTdngc2023_sisio2} we show 
temperature distribution functions $P(a,T)$ for pure Si nanoparticles
and Si/SiO$_2$ nanoparticles of various sizes.
In each case we show $P(a,T)$ for both neutral and charged grains.


The 2.4--45$\um$ spectrum has been obtained by Verstraete et al.\ (2001)
using the Infrared Space Observatory, using apertures ranging from
$14^{\prime\prime}\times 20^{\prime\prime}$ (for 2.4--12$\mu$m) to
$20^{\prime\prime}\times 33^{\prime\prime}$ (for 29--45$\mu$m), 
at a position $60^{\prime\prime}$ S of HD 37903.
Far-infrared photometry with a 37$^{\prime\prime}$ beam 
centered 60$^{\prime\prime}$ S of HD 37903 has
been obtained from the Kuiper Airborne Observatory 
by Harvey et al.\ (1980).
The observed IR spectrum is shown in Figure \ref{fig:emsn_ngc2023_si}.
Proposed ERE carriers must not produce IR emission in excess of what
is observed.

The PDR in NGC 2023 is optically thick to the 
illuminating starlight. 
Allowing for forward scattering, we estimate that the effective
attenuation cross section per H nucleon for 3--11 eV photons is
$\sim 0.5\times10^{-21}\cm^2 \,{\rm H}^{-1}$, 
and we approximate the PDR as a slab of 
optically-thin dust and gas with 
$N_{\rm H} \approx 2\times10^{21} {\rm cm}^{-2}$. 
We further assume a limb-brightening factor 
$1/\cos\theta\approx 2$, where $\theta$ is the angle 
between the slab normal and our line-of-sight
(our final results are insensitive to the adopted 
values of $N_{\rm H}$ and $1/\cos\theta$; see \S\ref{sec:discussion}). 
With the above assumptions, $j_\lambda$, the power radiated 
per H nucleon per unit solid angle per unit wavelength $\lambda$ 
(see eq.[\ref{eq:j_lambda}]) is related to the observed intensity
$I_\lambda$ by
\begin{equation}\label{eq:I_lambda}
I_\lambda = \frac{N_\rmH}{\cos\theta}j_\lambda 10^{-0.4A_\lambda}
\approx
4\times10^{21}\cm^{-2} j_\lambda 10^{-0.4A_\lambda}
\end{equation}
where $A_\lambda$ is the extinction between the
point of emission and the observer.

The foreground extinction $A_\lambda$ is uncertain.  
The line-of-sight to HD 37903 has $E(B-V)\approx 0.35$ and
$R_V\equiv A_V/E(B-V)=4.1$ (Cardelli, Clayton, \& Mathis 1989), 
corresponding to $A_{\rm 0.68\mu m} \approx 1.2$.
For the emission bar 80$^{\prime\prime}$ S of HD 37903, 
an extinction $A_{\rm 0.68\mu m}\approx 3.2$ has been 
determined from the relative strengths
of K band and far-red H$_2$ emission lines (Draine \& Bertoldi 2000).
For purposes of discussion, we will assume that
the extinction to the reflection region 60$^{\prime\prime}$
S of HD 37903 (where the ERE has been measured) is likely to
be in the range $1.2 \ltsim A_{\rm 0.68\mu m} \ltsim 3.2$.

To estimate $I^{\rm ERE}$ ($\equiv\int I_\lambda^{\rm ERE} d\lambda$)
60$^{\prime\prime}$ S of HD37903, we take
(1) the measured surface brightness profile from 
   Witt, Schild, \& Kraiman (1984, Figure 7);
(2) $I_{0.68\mu{\rm m}}/I_{0.62\mu{\rm m}}
   = 10^{0.06}F_{0.68\mu{\rm m}}^\star/F_{0.62\mu{\rm m}}^\star=0.91$
   $80^{\prime\prime}$ S of HD37903,
	where $F_\lambda^\star$ is the observed stellar flux
	(Witt \& Schild 1988, Table 1);
(3) the ERE spectral profiles 62$^{\prime\prime}$ and 84$^{\prime\prime}$
   ENE of HD37903 (Witt \& Boroson 1990, Figure 4).\footnote{%
	The ERE spectrum
	for 60$^{\prime\prime}$ S peaks at $6800\Angstrom$ 
	(Witt \& Schild 1988) as do the spectra
	62$^{\prime\prime}$ and 
	84$^{\prime\prime}$
	ENE of HD37903.
	}
From this we estimate 
$I^{\rm ERE}\approx 9\times10^{-5}\erg\cm^{-2}\s^{-1}\sr^{-1}$
60$^{\prime\prime}$ S of HD37903.
The ERE emissivity per H nucleon is
\begin{equation}
\label{eq:ERE_observed}
j^{\rm ERE} \equiv \int j_\lambda^{\rm ERE} d\lambda =
\frac{\int I_\lambda^{\rm ERE}10^{0.4A_\lambda}d\lambda}
{N_\rmH/\cos\theta }
\approx 2.3 \times 10^{-26} 10^{0.4A_{0.68\mu{\rm m}}}
\erg \s^{-1} \sr^{-1} {\rm H}^{-1} ~~~.
\end{equation}
Proposed ERE carriers must provide this level of emission.

\subsubsection{Si Nanoparticles\label{sec:pure_si}}

We first calculate the IR emission spectra for 
crystalline Si grains illuminated by radiation from HD 37903
with intensity $\chi=2000$. 
As shown in Figure \ref{fig:PTdngc2023_si},
the radiation field is so strong that neutral Si grains with 
$a\gtsim 30$\AA\ attain a steady-state temperature $\approx 820\K$,
while $a\gtsim 30$\AA\ charged Si grains are heated to $\simlt 200\K$.
The single-photon heating effect (Greenberg 1968) is evident
for $a_{\rm Si}=10\Angstrom$.
In Figure \ref{fig:emsn_ngc2023_si} we present the predicted spectra for
$\zsi=12$~ppm in Si nanoparticles\footnote{%
	As discussed in \S\ref{sec:dism}, we estimate that
	the observed ERE from the diffuse ISM would require 
	$\zsi\geq12$~ppm in Si SNPs,
	or $\zsi\geq15$~ppm in Si/SiO$_2$ SNPs.
	}
with radii $a=10, 20, 30, 40, 50\Angstrom$.
Grains with $a\ltsim 100$\AA\ are in the Rayleigh limit for
$\lambda >912$\AA, and 
thus their ``equilibrium'' temperatures are independent of $a$; 
therefore, for a given nanoparticle abundance 
$\zsi$, the resulting IR intensity is almost 
independent of grain size.

Although the fundamental lattice vibration of crystalline 
Si has no dipole moment and thus is IR inactive (i.e., 
there is no interaction between a single phonon and IR radiation), 
multi-phonon processes produce prominent bands at 6.91, 7.03, 
7.68, 8.9, 11.2, 13.5, 14.5, 16.4, and 17.9$\mu$m (Johnson 1959). 
As seen in Figure \ref{fig:emsn_ngc2023_si}a, for $\zsi=12$~ppm the 
$16.4\um$ emission feature is calculated to be
about 15 times stronger than the observed spectrum if the grains
remain neutral.
In order not to exceed the observed spectrum near $16.4\um$, 
NGC 2023 must have $\zsi\ltsim 0.8$~ppm in neutral Si SNPs -- 
less than 1/15 of the
abundance of SNPs to explain the ERE in the diffuse ISM.
If, however, the pure Si particles have charge +1, then the
thermal emission from the hole results in an emission continuum
peaking near 15--20$\micron$.
We see from Figure \ref{fig:emsn_ngc2023_si} that an abundance
$\zsi=12$~ppm in charged SNPs would result in emission which would
exceed the observed spectrum by about a factor $\sim10$ near 20$\micron$.
Thus we obtain an upper limit $\zsi\ltsim 1.5$~ppm in pure Si SNPs 
(all charge states) in NGC 2023.

Let $\etapl$ be the photoluminescence efficiency,
$\ratesi$ be the UV/visual photon absorption rate per Si
(in the 912--5500\AA\ wavelength range),
%$\nuabs$ be the mean absorbed photon energy, 
and $\nuere$ be the mean energy of ERE photons.
Since the NGC 2023 ERE peaks at $\sim 6800$\AA\ 
(Witt \& Boroson 1990), we take $\nuere \approx 1.8$eV.
Illuminated by starlight of $T_{\rm eff}$=22,000K,
we calculate
${\rm \ratesi \approx 7.43\times 10^{-6}(\chi/2000)\s^{-1}\ Si^{-1}}$
for neutral Si nanoparticles.
The ERE emissivity per H is then 
$j^{\rm ERE} = \nuere \ratesi \etapl \zsi /4\pi \approx 
1.70\times 10^{-18}\ \etapl \zsi \ {\rm erg\ s^{-1}\ sr^{-1}\ H^{-1}}$.
From eq. (\ref{eq:ERE_observed}) we thus require
$\etapl \zsi \approx 1.35\times10^{-8} 10^{0.4A_{\rm 0.68\mu m}}$.
Correspondingly, for charged Si particles, we obtain
${\rm \ratesi \approx 7.04\times 10^{-6}(\chi/2000)\s^{-1}\ Si^{-1}}$,
$j^{\rm ERE} \approx 1.62\times 10^{-18}\ \etapl \zsi \ 
{\rm erg\ s^{-1}\ sr^{-1}\ H^{-1}}$,
and $\etapl \zsi \approx 1.42\times10^{-8} 10^{0.4A_{\rm 0.68\mu m}}$.
Results are summarized in Table \ref{tab:rst}.

Since we have seen that $\zsi\ltsim 0.8$~ppm, it follows that pure Si
neutral nanoparticles could produce the observed ERE only if the
photoluminescence efficiency 
$\etapl \gtsim 0.017\times10^{0.4A_{0.68\um}}$.
For charged Si grains ($\zsi\ltsim 1.5$~ppm), 
$\etapl \gtsim 0.010\times10^{0.4A_{0.68\um}}$ is required. 
However, according to Smith \& Witt (2001),
those charged grains will not emit ERE, i.e., $\etapl \approx 0$. 

\begin{figure}[h]
\begin{center}
\epsfig{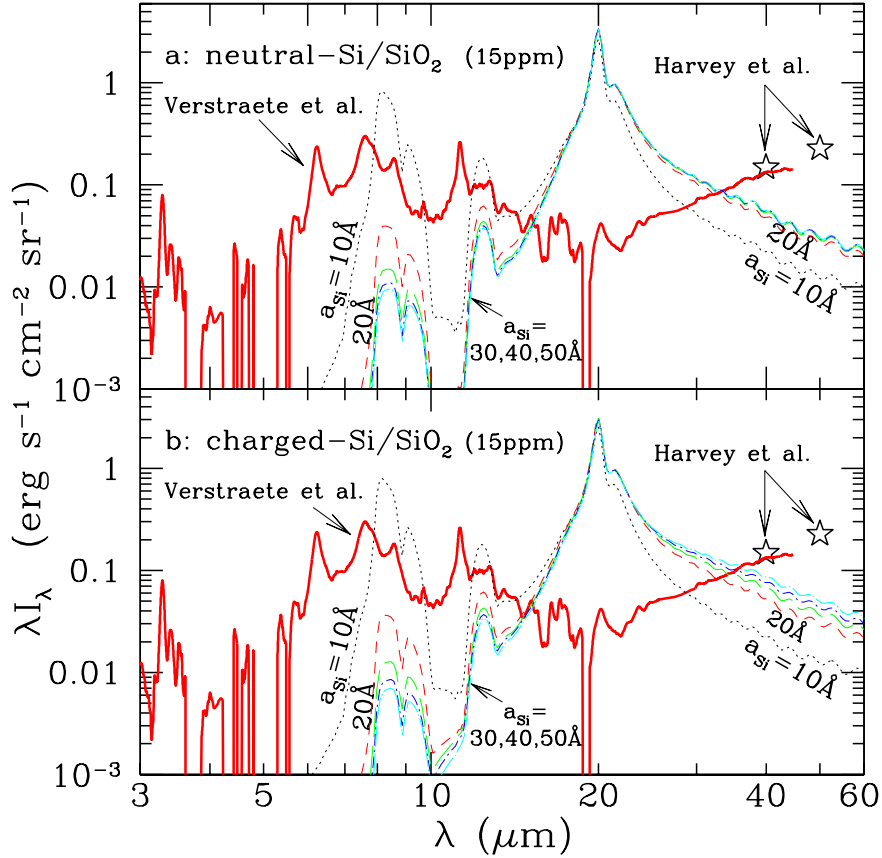}
\end{center}\vspace*{-2em}
\caption{
        \footnotesize
        \label{fig:emsn_ngc2023_sisio2}
	As in Figure \ref{fig:emsn_ngc2023_si}, but for
	Si/SiO$_2$ SNPs, for various values of the Si core radius
	$a_{\rm Si}=0.8a$.
        Upper panel (a) -- Model spectra calculated
        for neutral Si/SiO$_2$ grains containing Si/H=15~ppm,
	with radii $a_{\rm Si}=10,20,30,40,50\Angstrom$ (see text).
        Spectra for $a_{\rm Si}=20,30,40,50$\AA\ are 
	nearly indistinguishable. 
        Lower panel (b) -- As in (a), but for 
        charged Si/SiO$_2$ grains.
        The ripple features at $\lambda >30\mu$m are caused
        by the SiO$_2$ optical constants adopted from
        Henning \& Mutschke (1998).  
        }
\end{figure}

\subsubsection{Si/SiO$_2$ Nanoparticles\label{sec:si_sio2}} 

It seems unlikely that {\it pure} Si nanoparticles 
would exist in interstellar space without some degree 
of oxidation.
Si nanocrystals in air spontaneously form an overlayer of 
SiO$_2$ which is usually $\sim 10$\AA\ thick 
(Ledoux et al.\ 1998). Witt et al.\ (1998) propose that SNPs
in the form of Si core-SiO$_2$ mantle structure 
are a product of the initial dust formation process 
in oxygen-rich stellar mass outflows. 
Laboratory studies show that the oxidation of the Si core 
is a self-limiting process and, in the final stage of oxidation, 
the thickness of the oxide layer represents about 10\% of 
the total particle diameter (Ledoux et al.\ 2000).

We now calculate the IR emission spectra for 
Si core-SiO$_2$ mantle particles (both neutral and charged). 
Following Ledoux et al.\ (2000), 
we assume $a_{\rm Si}=0.8a$ where $a_{\rm Si}$ and $a$
are respectively the radii of the Si core and the Si/SiO$_2$ entire grain.
In Figure \ref{fig:emsn_ngc2023_sisio2} we show theoretical spectra for
$a_{\rm Si}=10, 20, 30, 40, 50$\AA\ with $\zsi = 15$~ppm 
(estimated in \S\ref{sec:dism} for Si/SiO$_2$ grains to
account for the ERE in the diffuse ISM)
together with the observed spectrum for NGC 2023. 
As illustrated in Figure \ref{fig:emsn_ngc2023_sisio2}
the Si-O modes produce strong features at 
20$\mu$m,
%(transverse vibration), 
12.5$\mu$m 
%(symmetric pumping vibration), 
and 9.1$\mu$m.
%(asymmetric valency vibration).
The most conspicuous feature is the strong and broad 20$\mu$m
band which, for $\zsi=15$~ppm,
is $\sim$80 times stronger than the observational data. 
To depress the 20$\mu$m emission feature to a level
not in contradiction with the observational spectrum, 
NGC 2023 must have $\zsi \ltsim 0.2$~ppm in 
SiO$_2$-coated Si grains (all charge states).

The 912--5500\AA\ photon absorption rate for 
these Si/SiO$_2$ neutral SNPs is 
$\ratesi \approx 6.19\times 10^{-6}
(\chi/2000)\ \s^{-1}\ {\rm Si}^{-1}$.
The ERE emissivity per H is then
$\nuere \ratesi \etapl \zsi /4\pi \approx 
1.42\times 10^{-18}\ \etapl \zsi \ {\rm erg\ s^{-1}\ sr^{-1}\ H^{-1}}$.
From eq. (\ref{eq:ERE_observed}) we thus require
$\etapl \zsi \approx 1.62\times10^{-8} 10^{0.4A_{\rm 0.68\mu m}}$.
Since $\zsi \ltsim 0.2$~ppm the {\it observed} ERE intensity 
in NGC 2023 (eq.[\ref{eq:ERE_observed}]) requires
a photoluminescence efficiency 
$\etapl \simgt 0.081\times 10^{0.4 A_{\rm 0.68\mu m}}$ if
due to Si core-SiO$_2$ mantle neutral nanoparticles.
Similarly, we obtain for Si/SiO$_2$ charged SNPs 
$\ratesi \approx 6.00\times 10^{-6}
(\chi/2000)\ \s^{-1}\ {\rm Si}^{-1}$,
$j^{\rm ERE} \approx 1.37\times 10^{-18}\ \etapl \zsi \
{\rm erg\ s^{-1}\ sr^{-1}\ H^{-1}}$,
$\etapl \zsi \approx 1.68\times10^{-8} 10^{0.4A_{\rm 0.68\mu m}}$,
and $\etapl \simgt 0.084\times 10^{0.4 A_{\rm 0.68\mu m}}$.
Again, according to Smith \& Witt (2001),
those charged grains will have $\etapl \approx 0$. 
Results are also summarized in Table \ref{tab:rst}.

\subsection{Diffuse Interstellar Medium\label{sec:dism}}

Using Pioneer 10 and 11 observations, Gordon et al.\ (1998) estimated
the ERE emissivity per H nucleon of the high Galactic latitude (HGL) 
diffuse ISM to be $j^{\rm ERE} \approx 1.4\times 10^{-26} \ 
\erg \s^{-1} \sr^{-1} {\rm H}^{-1}$.
Szomoru \& Guhathakurta (1998) obtained $\langle I^{\rm ERE}\rangle
= 1.2\times10^{-5}\erg\s^{-1}\sr^{-1}{\rm cm}^{-2}$ toward 3 cirrus clouds
with $\langle N_{\rm H}\rangle=7.3\times10^{20}\cm^{-2}$, giving
$j^{\rm ERE}\approx 1.6\times10^{-26}\erg\s^{-1}\sr^{-1}\rmH^{-1}$.
Evidently the ERE carrier must provide an emissivity
$j^{\rm ERE}\approx 1.5\times10^{-26}\erg\s^{-1}\sr^{-1}\rmH^{-1}$
in the diffuse ISM.
The ERE appears to peak at $\sim$6000$\Angstrom$ 
(Szomoru \& Guhathakurta 1998);
if due to SNPs, this would indicate $2a\approx 28\Angstrom$
(Ledoux et al.\ 2000), or $a_{\rm Si}\approx 0.8a\approx 11\Angstrom$.

\subsubsection{Si Nanoparticles\label{sec:dismsi}}

\begin{figure}[h]
\begin{center}
\epsfig{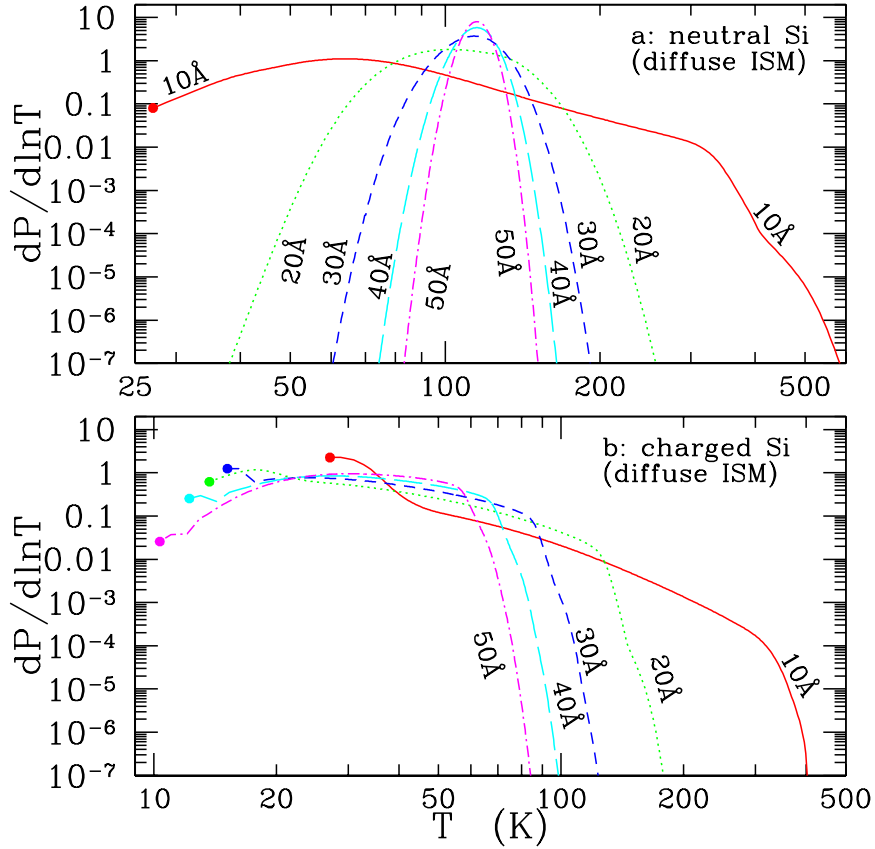}
\end{center}\vspace*{-2em}
\caption{
        \footnotesize
        \label{fig:pdlnT_dism_si}
        Temperature distribution functions 
        for pure Si grains illuminated by the local
	interstellar radiation field
	(Mathis, Mezger, \& Panagia 1983).
	Upper panel: neutral Si grains.
	Lowel panel: charged Si grains.
        For 10\AA\ (upper panel), 10--50\AA\ (lower panel)
        a dot indicates the first vibrationally-excited state 
	(see Draine \& Li 2001). 
        }
\end{figure}

\begin{figure}[h]
\begin{center}
\epsfig{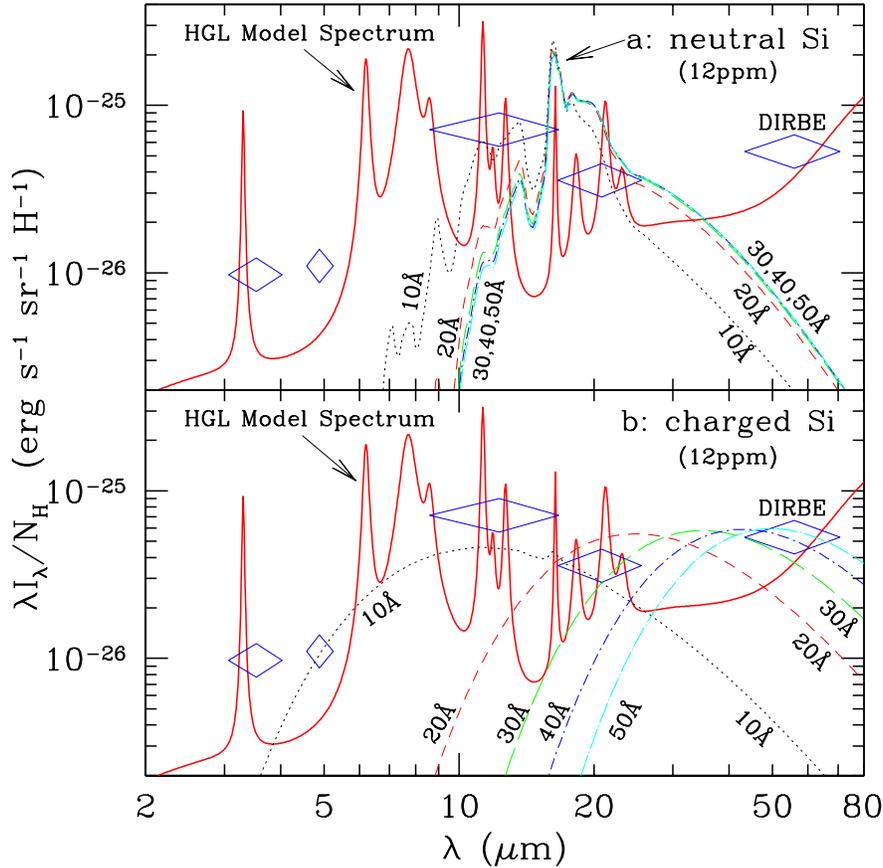}
\end{center}\vspace*{-2em}
\caption{
        \footnotesize
        \label{fig:emsn_dism_si}
        Predicted IR emission spectra for pure Si SNPs
	with $\zsi=12$~ppm
        in the diffuse ISM,
	for (a) neutral Si grains, and
	(b) $Z=+1$ Si grains.
	Curves are labelled by grain radius.
        Also plotted are the DIRBE photometry (diamonds; Arendt et al.\ 1998) 
        and the HGL model spectrum of Li \& Draine (2001).
        }
\end{figure}

The 912--5500\AA\ photon absorption rates $\ratesi$ for 
pure Si neutral and charged grains are 
$\approx 4.94\times 10^{-9} (\chi/1.23) \s^{-1}\ {\rm Si}^{-1}$, 
and $\approx 4.68\times 10^{-9} (\chi/1.23) \s^{-1}\ {\rm Si}^{-1}$, 
respectively. Thus
$j^{\rm ERE}\approx 1.32\times 10^{-21}(\chi/1.23)
\etapl \zsi \ {\rm erg\ s^{-1}\ sr^{-1}\ H^{-1}}$,
and
$j^{\rm ERE}\approx 1.25\times 10^{-21}(\chi/1.23)
\etapl \zsi \ {\rm erg\ s^{-1}\ sr^{-1}\ H^{-1}}$,
respectively, for neutral and charged Si grains. 
Therefore, since $\etapl \leq 100$\%, we require
$\zsi \gtsim 12$~ppm ($\gtsim 33\%$ of $\sisun$) 
for both neutral and charged Si grains. 
In Figure \ref{fig:pdlnT_dism_si} 
we plot the temperature probability
distribution functions for these grains, and
IR emission spectra 
are shown in Figure \ref{fig:emsn_dism_si}.

The DIRBE $25\micron$ photometry is inconsistent
with $\zsi\approx 12$~ppm in pure Si neutral nanoparticles, 
with an upper limit $\zsi\ltsim 4$~ppm. 
For charged Si grains, the predicted IR spectra are not in
conflict with the DIRBE photometry. However, we note again
that, according to Smith \& Witt (2001),
those charged grains will not emit ERE.
Results are summarized in Table \ref{tab:rst}.

\subsubsection{Si/SiO$_2$ Nanoparticles\label{sec:dismsio2}}
Similarly, we have also considered Si core-SiO$_2$ mantle grains
with $a_{\rm si}=0.8a$, assuming the Si core to be either
neutral or have charge $+1$. 

\begin{figure}[h]
\begin{center}		% figure 9
\epsfig{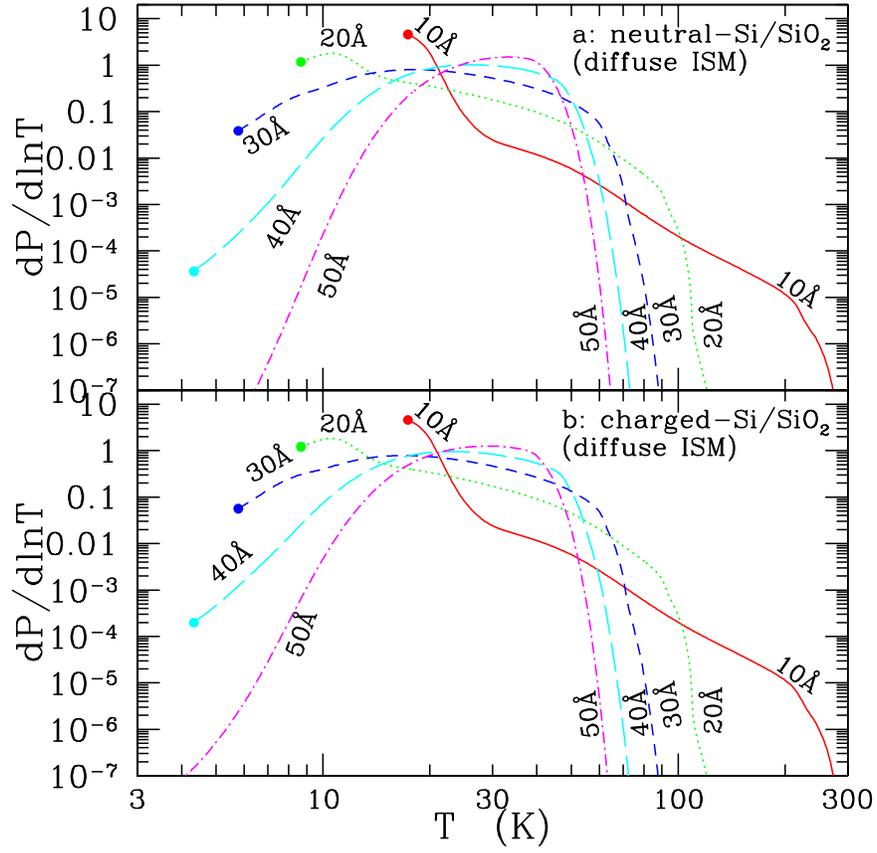}
\end{center}\vspace*{-2em}
\caption{
        \footnotesize
        \label{fig:pdlnT_dism_sisio2}
	Temperature distribution functions for Si/SiO$_2$ spheres
	of various radii, with the Si cores assumed to 
        be (a) neutral and (b) charged with $Z=+1$,
	for grains illuminated by the local interstellar 
        radiation field (Mathis, Mezger, \& Panagia 1983).
	Curves are labelled by the Si core radius $a_{\rm Si}$
        ($\approx 0.8 a$ where $a$ is the Si/SiO$_2$ radius).
        For $a_{\rm Si}=$10, 20, 30, 40\AA\ 
        a dot indicates the first excited state (see Draine \& Li 2001). 
        }
\end{figure}

\begin{figure}[h]
\begin{center}		% figure 10
\epsfig{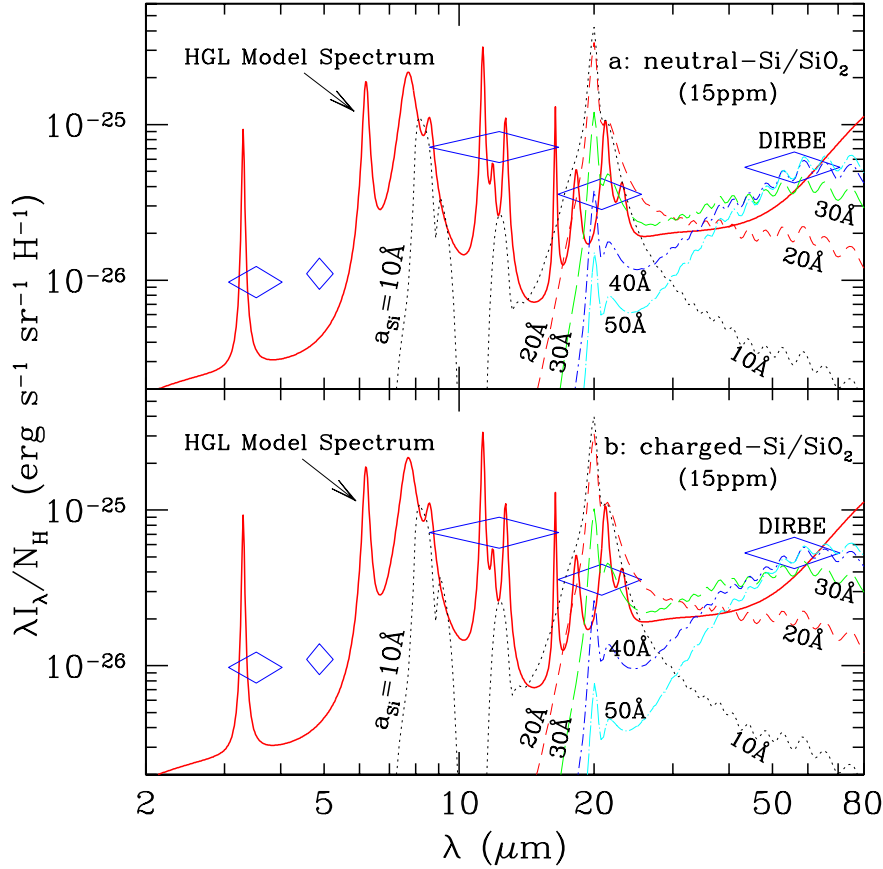}
\end{center}\vspace*{-2em}
\caption{
        \footnotesize
        \label{fig:emsn_dism_sisio2}
	As in Fig. \ref{fig:emsn_dism_si}, but for
        Si/SiO$_2$ SNPs with $\zsi=15$~ppm.
        }
\end{figure}

The 912--5500\AA\ photon absorption rates $\ratesi$ for 
Si/SiO$_2$ neutral and charged grains are 
$\approx 3.87\times 10^{-9} (\chi/1.23) \s^{-1}\ {\rm Si}^{-1}$, 
and $\approx 3.77\times 10^{-9} (\chi/1.23) \s^{-1}\ {\rm Si}^{-1}$, 
respectively. Thus
$j^{\rm ERE}\approx 1.04\times 10^{-21}(\chi/1.23)
\etapl \zsi \ {\rm erg\ s^{-1}\ sr^{-1}\ H^{-1}}$,
and
$j^{\rm ERE}\approx 1.00\times 10^{-21}(\chi/1.23)
\etapl \zsi \ {\rm erg\ s^{-1}\ sr^{-1}\ H^{-1}}$,
respectively, for neutral and charged Si/SiO$_2$ grains. 
Therefore, since $\etapl \leq 100$\%, we require
$\zsi \gtsim 15$~ppm ($\gtsim 42\%$ of $\sisun$) for Si/SiO$_2$ grains. 
In Figure \ref{fig:pdlnT_dism_sisio2} 
we plot the temperature probability
distribution functions for these grains, and
IR emission spectra are shown in Figure \ref{fig:emsn_dism_sisio2}.
Results are also summarized in Table \ref{tab:rst}.

Evidently, the DIRBE photometry rules out 
$\zsi\approx 15$~ppm in Si/SiO$_2$ nanoparticles 
(both neutral and charged) with radii $a_{\rm Si}\ltsim35\Angstrom$.
The DIRBE photometry places an upper limit
$\zsi\ltsim 2$~ppm on the abundance of the 
$a_{\rm Si}\approx 11\Angstrom$ particle
size which appears to be required by the observed wavelength of peak
emission in the diffuse ISM.
Thus, we conclude that free-flying SNPs can account for no more than
13\% of the observed ERE from the diffuse ISM.

SIRTF will be able to obtain 
diffuse interstellar cloud emission spectra 
which should either detect the 16.4 or 20$\micron$
spectral features, or place much stronger constraints 
on the abundance of SNPs.
%The required $\zsi$ could be slightly reduced 
%for nonspherical grains 
%(for example, the UV absorption cross section 
% of a 5:1 prolate Si/SiO$_2$ grain is $\sim$10\% 
%larger than an equal-volume sphere 
%[see Figure \ref{fig:cabs}]),
%or if the interstellar radiation field is stronger than estimated by
%Mathis, Mezger, \& Panagia (1983), but these effects are minor.

\section{Discussion\label{sec:discussion}}

Si/SiO$_2$ core-mantle grains are heated primarily by absorption
by the Si core (the photoluminescence of which is responsible for the
ERE), since the SiO$_2$ mantle is almost nonabsorptive for
$0.12\ltsim\lambda\ltsim 4\um$; the core-mantle grains are
cooled primarily by the SiO$_2$ mantle, since $m^{\prime\prime}$
is so small for Si in the infrared (see Figure \ref{fig:opct}).
Therefore, increasing the amount of SiO$_2$ present (while holding the
pure Si cores constant) will lead to only small changes in the IR
emission resulting from the small decrease in average grain 
temperature (e.g., Si/SiO$_2$ grains with equal numbers of Si atoms
in the core and in the mantle [$a_{\rm Si}\approx 0.668a$]
in NGC 2023 [$\chi$=2000] have a steady-state temperature 
$\approx 85$K in comparison with $\approx$96K for grains with
$a_{\rm Si}=0.8a$). The limits on $\etapl$ will be only slightly 
affected by increasing the thickness of the SiO$_2$ mantles.

\begin{figure}[h]
\begin{center}		% figure 11
\epsfig{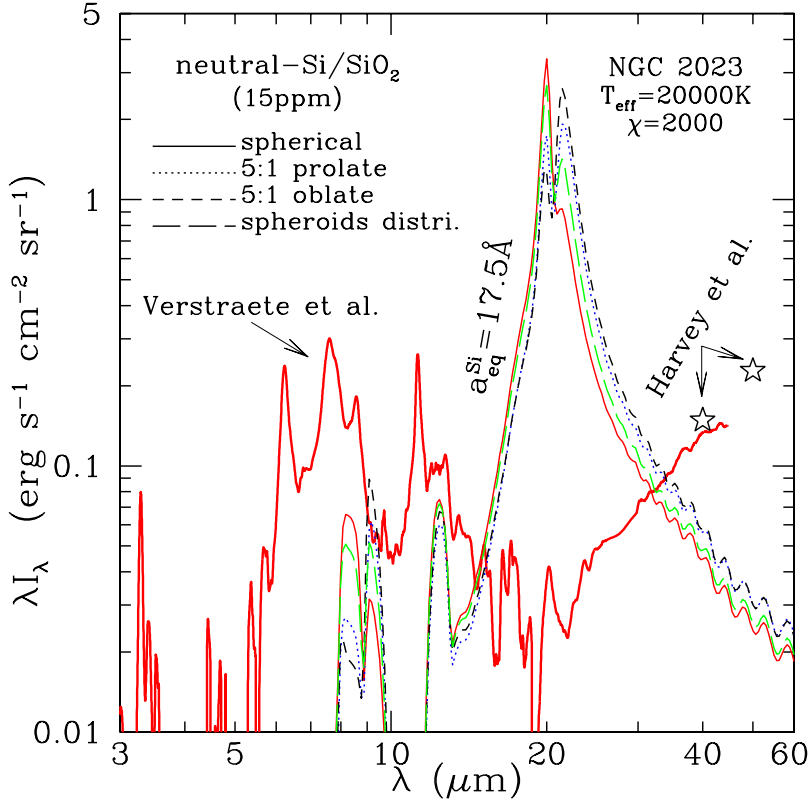}
\end{center}\vspace*{-2em}
\caption{
        \footnotesize
        \label{fig:shape}
        Grain shape effects on the model IR emission spectra
        calculated from spherical grains (thin solid line),
        5:1 prolate grains (dotted line), 
        5:1 oblate grains (short-dashed line),
        and grains with a distribution of spheroidal shapes 
        (long-dashed line) illuminated by the NGC 2023 radiation field.
        All grains are taken to 
        (1) be neutral; 
        (2) have an equivalent sphere-volume Si core size 
        $a_{\rm eq}^{\rm Si}=17.5$\AA\ (we choose 17.5\AA\ because
        Smith \& Witt [2001] found that the observed
        ERE in the diffuse ISM can be best explained by SNPs with 
        $a_{\rm Si}=17.5$\AA);
        (3) have the Si core taking over 51\% of the total grain volume.
        Also plotted are the ISO SWS spectrum (heavy solid line;
        Verstraete et al.\ 2001)
        and the KAO photometry (stars; Harvey et al.\ 1980). 
	It is evident that the grain shape effect does not
        affect our conclusions. 
        }
\end{figure}

In \S\ref{sec:results} the grains are modelled as spherical.
To investigate the sensitivity to shape variations,
we have also calculated the IR emission spectra for 
(1) 5:1 prolate grains;
(2) 5:1 oblate grains;
(3) grains with a distribution of spheroidal shapes with 
$dP/dL_{\parallel} = 12 L_{\parallel}[1-L_{\parallel}]^2$ 
(Ossenkopf, Henning, \& Mathis 1992) where $L_{\parallel}$ is the 
so-called ``depolarization factor'' parallel to the grain 
symmetry axis (for spheres $L_{\parallel}$=1/3); this
shape distribution peaks at spheres and drops to zero for 
the extreme cases $L_{\parallel}\rightarrow 0$ (infinitely thin needles)
or $L_{\parallel}\rightarrow 1$ (infinitely flattened pancake). 
For core-mantle grains, we assume confocal geometry,
with the above $dP/dL_{\parallel}$ applying to the outer surface.
For illustration, we present in Figure \ref{fig:shape} the IR spectra
calculated for neutral Si/SiO$_2$ grains 
illuminated by the NGC 2023 radiation field. 
It is clear that the shape effects are minor:
the major change is a slight shift of the peak wavelength 
of the 20$\mu$m Si-O feature.

In a reflection nebula such as NGC 2023, 
the IR surface brightness due to SNPs 
$\int I_\lambda d\lambda \propto N_{\rm H}\chi \zsi/\cos\theta$.
While the fraction of the IR power radiated in the $20\um$ SiO$_2$ feature
(or the $16.4\um$ Si feature in the case of pure Si nanoparticles)
depends on $\chi$ (through the grain temperature), this dependence
is relatively weak for modest changes in $\chi$.
Therefore the
upper limits on the 16.4 and 20$\micron$ features in NGC 2023
give upper limits on
$N_{\rm H}\chi \zsi/\cos\theta$;
these limits, plus the observed ERE 
$\propto \etapl N_{\rm H}\chi \zsi/\cos\theta$
then give lower limits on $\etapl$,
independent of the actual values of $N_{\rm H}$ and
$\cos\theta$, and only weakly dependent
on the value of $\chi$.

One considerable uncertainty in our analysis concerns 
the UV absorption properties of nano-Si material.
In the literature, it is often stated that, as a result
of quantum confinement, the 
optical-UV absorption of
%dielectric functions of
nano-Si is substantially reduced in comparison with
those of bulk-Si. If this is true, the SNP model
would require an even larger Si abundance to account for the ERE, 
and the limitations
placed on this model would be even more severe. 

Another uncertainty concerns the IR emission properties 
of nano-Si crystals and nano-SiO$_2$, which have not yet 
been experimentally investigated. In the present work, our results are 
mainly based on the IR emissivities of bulk Si and bulk SiO$_2$. 
Nano-materials are expected to be more IR-active than
their bulk counterparts (due to the symmetry-breaking surfaces)
and therefore the detailed emission spectrum will be modified.
However, the starlight energy absorbed by the grains has to
be reradiated in the infrared, so the absence of detected spectral
features implies upper limits on the SNP abundance.

Our analysis of NGC 2023 required the value of
the foreground extinction $A_\lambda$ to ``deredden''
the ERE measurements.
Unfortunately, $A_\lambda$ has not been directly measured
at the location where the ERE observations have been made.
It would be extremely valuable to have ERE measurements made at a
position where a reliable extinction has been determined -- such
as the emission bar 80$^{\prime\prime}$ S of HD 39037 -- or to have
both IR and far-red observations of $\HH$ emission (allowing $A_\lambda$
to be determined) at a location where
the ERE has been measured.

Our modelling has assumed isolated ``free-flying'' SNPs;
we have demonstrated that existing DIRBE photometry for the diffuse ISM
rules out the hypothesis that the observed ERE is due to such particles.
Since isolated SNPs are ruled out, it is important to note that
in the laboratory, photoluminescence is observed for
SNPs located on a substrate (Ledoux et al.\ 2000).
An SNP located on the surface of a large interstellar grain
would have its temperature -- and therefore its infrared emission --
determined by the temperature of the ``host'' grain.
In the case of NGC 2023, the radiation field is so intense that
the ``host'' temperature would probably be similar to the temperature of
the Si/SiO$_2$ SNPs, so the 20$\micron$ SiO$_2$ emission feature should
still be prominent, though its strength could be increased or decreased,
depending on the optical properties of the ``host'' grain.
If the SNPs are pure Si, however, the 16.4$\micron$ emissivity would
be reduced if the SNPs are attached to a host, since the
host grain would not be as hot as the Si SNPs, which are poor
thermal emitters.
In the diffuse ISM, the 16.4 and 20$\micron$
emission would be completely suppressed 
if the SNPs are attached to larger host grains, 
with temperatures $T\approx 15-20\K$. 
Thus SNPs attached to larger host grains may be a viable explanation for
the observed ERE.

The photoexcitation rate $\gamma$ for SNPs on the surfaces of
$a\gtsim 100\Angstrom$ grains would probably be lower than for
free-flying SNPs, which would in turn increase the abundance of Si
in SNPs required to explain the observed ERE.
Since 42\% or more of 
$\sisun$ is required even for free-flying SNPs, the Si abundance
demands would be difficult to accommodate if the SNPs are
attached to $a\gtsim100\Angstrom$ grains.

\begin{table}
{
%\scriptsize
\caption[]{ERE Emissivities of SNPs
	\label{tab:rst}}
\begin{tabular}{llll}
\hline \hline
\multicolumn{2}{c}{Item} & NGC 2023 & Diffuse ISM\\ 
\hline
Observational & $\langle h\nu\rangle_{\rm ERE}\ (\eV)$ & 1.8 & 2.1\\
\cline{2-4}
 ~   & $j^{\rm ERE}_{\rm obs}\ (\erg \s^{-1} \sr^{-1} {\rm H}^{-1})^{a}$ 
     & $2.3 \times 10^{-26} 10^{0.4A_{0.68\mu{\rm m}}}$
     & $1.5 \times 10^{-26}$\\
\hline
 & $\langle h\nu\rangle_{\rm abs}\ (\eV)^{b}$ & 9.24 & 9.25\\
\cline{2-4}
neutral Si & ${\rm \ratesi\ (\s^{-1}\ {\rm Si}^{-1})}^{c}$
   & $7.43\times 10^{-6}(\chi/2000)$
   & $4.94\times 10^{-9}(\chi/1.23)$\\
\cline{2-4}
 ~	& $j^{\rm ERE}_{\rm mod}\ ({\rm erg\ s^{-1}\ sr^{-1}\ H^{-1}})^{d}$
  & $1.70\times 10^{-18}\ \etapl \zsi$
  & $1.32\times 10^{-21}\ \etapl \zsi$\\
\hline
 & $\langle h\nu\rangle_{\rm abs}\ (\eV)$ & 9.30 & 9.29\\
\cline{2-4}
charged Si & ${\rm \ratesi\ (\s^{-1}\ {\rm Si}^{-1})}$
   & $7.04\times 10^{-6}(\chi/2000)$
   & $4.68\times 10^{-9}(\chi/1.23)$\\
\cline{2-4}
~ & $j^{\rm ERE}_{\rm mod}\ ({\rm erg\ s^{-1}\ sr^{-1}\ H^{-1}})$
  & $1.62\times 10^{-18}\ \etapl \zsi$
  & $1.25\times 10^{-21}\ \etapl \zsi$\\
\hline
 & $\langle h\nu\rangle_{\rm abs}\ (\eV)$ & 8.59 & 8.59\\
\cline{2-4}
neutral Si/SiO$_2$ & ${\rm \ratesi\ (\s^{-1}\ {\rm Si}^{-1})}$
   & $6.19\times 10^{-6}(\chi/2000)$
   & $3.87\times 10^{-9}(\chi/1.23)$\\
\cline{2-4} 
  & $j^{\rm ERE}_{\rm mod}\ ({\rm erg\ s^{-1}\ sr^{-1}\ H^{-1}})$
  & $1.42\times 10^{-18}\ \etapl \zsi$
  & $1.04\times 10^{-21}\ \etapl \zsi$\\
\hline
 & $\langle h\nu\rangle_{\rm abs}\ (\eV)$ & 8.65 & 8.62\\
\cline{2-4}
charged-Si/SiO$_2$ & ${\rm \ratesi\ (\s^{-1}\ {\rm Si}^{-1})}$
   & $6.00\times 10^{-6}(\chi/2000)$
   & $3.77\times 10^{-9}(\chi/1.23)$\\
\cline{2-4}
~ & $j^{\rm ERE}_{\rm mod}\ ({\rm erg\ s^{-1}\ sr^{-1}\ H^{-1}})$
  & $1.37\times 10^{-18}\ \etapl \zsi$
  & $1.00\times 10^{-21}\ \etapl \zsi$\\
\hline
\end{tabular} \\[2mm]
\tablenotetext{a}{For NGC 2023, $j_{\rm obs}^{\rm ERE}$ is estimated 
	assuming a UV-illuminated slab with $N_{\rm H}=2\times10^{21}\cm^{-2}$
	with limb-brightening $1/\cos\theta=2$ (see text).}
\tablenotetext{b}{$\langle h\nu \rangle_{\rm abs} \equiv
\int_{912\AA}^{5500\AA} C_{\rm abs}(h\nu) c u_\nu d\nu/
\int_{912\AA}^{5500\AA} C_{\rm abs}(h\nu)(cu_\nu/h\nu) d\nu$
where $cu_\nu$ is the starlight intensity.} 
\tablenotetext{c}{$\ratesi \equiv N_{\rm Si}^{-1}
\int_{912\AA}^{5500\AA} C_{\rm abs}(h\nu)(cu_\nu/h\nu) d\nu$.} 
\tablenotetext{d}{$j^{\rm ERE}_{\rm mod} = \nuere \ratesi \etapl \zsi /4\pi$.}
}
\end{table}

%%%%%%%%%%%%%%%%%%%%%%%%%%%%%%%%%%%%%%%%%%%%%%%%%%%%%%%%%%%%%%%%%%%%%%%%%%%%%

\begin{table}
{
%\scriptsize
\footnotesize
\caption[]{IR Limits on Si/H (ppm) in Silicon Nanoparticles
	\label{tab:Zsi}}
\begin{tabular}{cccccc}
\hline \hline
 & & 
\multicolumn{2}{c}{neutral Si}
&\multicolumn{2}{c}{charged Si\tablenotemark{a}}\\
\cline{3-4}
\cline{5-6}
Region & Composition & Si/H (ppm) from & Si/H (ppm) from 
       & Si/H (ppm) from & Si/H (ppm) from\\ 
       & & observed IR emission & observed ERE intensity\tablenotemark{b}
       & observed IR emission & observed ERE intensity\tablenotemark{b}\\  
\hline
NGC 2023 & Pure Si & $<0.8$
         & $0.014~ \etapl^{-1}\times10^{0.4A_{0.68\mu{\rm m}}}$
         & $<1.5$ & $0.014~ \etapl^{-1}\times10^{0.4A_{0.68\mu{\rm m}}}$\\
	 & Si/SiO$_2$
         & $<0.2$&$0.016~\etapl^{-1} \times10^{0.4A_{0.68\mu{\rm m}}}$
         & $<0.2$&$0.017~\etapl^{-1} \times10^{0.4A_{0.68\mu{\rm m}}}$\\
\hline
Diffuse ISM & Pure Si&	$<4$&	$12~\etapl^{-1}$
                     &	$<12$&	$12~\etapl^{-1}$\\
	& Si/SiO$_2$ &	$<2$&	$15~\etapl^{-1}$
                     &	$<2$&	$15~\etapl^{-1}$\\
\hline
\end{tabular}\\[2mm]
\tablenotetext{a}{Note charged SNPs do not luminesce (i.e. $\etapl =0$).}
\tablenotetext{b}{$\etapl < 1$ is the photoluminescence efficiency.
	Lower limits are obtained by setting $\etapl = 1$.
	$A_{0.68\mu{\rm m}}$ is the 60$^{\prime\prime}$S NGC 2023 
        foreground extinction at 0.68$\mu$m (see text).}}
\end{table}

\section{Conclusions\label{sec:summary}}

Table \ref{tab:Zsi} summarizes our results for the abundances of SNPs:
upper limits based on nondetection of 16.4 and 20$\micron$ features,
and lower limits (since $\etapl \leq 1$) if neutral SNPs are to 
account for the observed ERE. Note charged SNPs are not able to 
emit ERE (Smith \& Witt 2001).

For pure (both neutral and charged) Si SNPs, nondetection of a
16.4$\mu$m emission feature in NGC 2023
implies $\etapl \gtsim 0.017\times 10^{0.4A_{0.68\um}}$.
Thus for $A_{0.68\um}=1.2$, pure neutral Si SNPs could account for
the observed ERE in NGC 2023 provided $\etapl\gtsim 5\%$, but if
$A_{0.68\um}=3.2$, 
$\etapl \gtsim 32\%$ would be required to account for the ERE.  
%While such a high value cannot be excluded,
We note in the laboratory high photoluminescence efficiencies 
are seen only when the Si surfaces are ``passivated'' 
by an oxide coating.

An oxide coating, however, will introduce a strong
emission feature at 20$\mu$m which is not seen in NGC 2023.
For Si/SiO$_2$ nanoparticles to explain the ERE, 
nondetection of the 20$\mu$m feature requires
a luminescence efficiency
$\etapl \gtsim 0.081\times10^{0.4A_{0.68\um}}$.
Photoluminescence efficiencies as high as 50\% have been reported 
for Si/SiO$_2$ nanoparticles (Wilson, Szajowski, \& Brus 1993), 
and in principle $\etapl$ could approach 100\%. 
However, if $A_{0.68\um}\gtsim 2.8$,
the ERE could not be due to Si/SiO$_2$ nanoparticles 
since the required $\etapl$ would then exceed 100\%.

The ERE emissivity of the diffuse ISM (Gordon et al.\ 1998;
Szomoru \& Guhathakurta 1998)
requires $\gtsim42\%\sisun$ in neutral Si/SiO$_2$ SNPs,
even for photoluminescence efficiency $\etapl\rightarrow100\%$.
Such a high abundance of SNPs
in the diffuse ISM is difficult to reconcile with the
evidence that a substantial fraction of interstellar Si is in fact
locked up in amorphous silicate grains 
(e.g., see Weingartner \& Draine 2001).
This difficulty is exacerbated if interstellar abundances are significantly
subsolar, as has been argued (see Snow \& Witt 1996).

We have calculated the  
IR emission which such particles would produce if they are free-flying.
Existing DIRBE 25$\micron$ photometry appears to already 
rule out such high abundances of SNPs, indicating that if SNPs are
responsible for the ERE from the diffuse ISM,
they must either be
in $a\gtsim 50\Angstrom$ clusters, or attached to larger ``host'' grains.
Future observations by SIRTF will be even more sensitive to
the abundance of free-flying SNPs in the diffuse interstellar medium.

In light of the controversy over the optical properties of
nano-Si material, we call for further experimental 
studies to determine the dielectric functions of nano-Si material
either directly from nano-Si crystals, 
or from porous-Si with the effects of voids and surface oxidation 
well determined and removed.

Finally, we also note that, if the surface dangling bonds 
of Si grains are passivated by H atoms and if they are present 
in interstellar space in quantities sufficient to account for 
the observed ERE, we would expect to see the Si-H 15.6$\mu$m wagging, 
11.6$\mu$m bending, and even 5$\mu$m stretching bands (Adachi 1999).

\acknowledgments
We especially thank A.N. Witt for invaluable comments and helpful
suggestions,
and L. Verstraete for 
providing us with the ISO spectrum of NGC 2023 in advance of
publication.
We also thank S. Adachi, N. Koshida, P. Maddalena, and 
V.G. Zubko for sending us their optical constants of 
bulk Si and/or nano-silicon; 
S. Adachi, L.J. Allamandola, J.M. Greenberg, Th. Henning, 
N. Koshida, D. Kovalev, P. Maddalena, L. Pavesi,
and K. Sellgren for helpful suggestions; and
R.H. Lupton for the availability of the SM plotting 
package. 
This research was supported in part by 
NASA grant NAG5-7030 and NSF grants AST-9619429 and AST-9988126.

\end{document}